\documentclass[10pt,a4paper,3p,preprint,number,sort&compress,times,english]{elsarticle}
\newif\ifdraft
\draftfalse

\usepackage[T1]{fontenc}
\usepackage[utf8]{inputenc}

\usepackage{babel}
\usepackage[colorlinks]{hyperref}
\usepackage[subrefformat=parens]{subcaption}
\usepackage{graphicx}
\usepackage{rotating}
\usepackage{multirow}
\usepackage{amscd}
\usepackage[centertags,reqno]{amsmath}
\usepackage{amssymb}
\usepackage{theorem}
\usepackage{paralist}
\usepackage{balance}
\usepackage{fancyhdr}

\makeatletter
\def\ps@pprintTitle{%
     \let\@oddhead\@empty
     \let\@evenhead\@empty
     \def\@oddfoot{\footnotesize\itshape
       r.\gitRevision\hfill\gitAuthorDate}%
     \let\@evenfoot\@oddfoot}
\makeatother

\allowdisplaybreaks

\newtheorem{definition}{Definition}
\newtheorem{proposition}{Proposition}
\newtheorem{theorem}{Theorem}

\newcommand{\inserttabd}[4]{%
   \begin{table*}[!tp]
      \begin{minipage}{\textwidth}
         \caption{#3}
         \label{#1}
         \centering
         \renewcommand{\arraystretch}{1.15}
         \begin{tabular}{#2}
            \hline
            #4
            \hline
         \end{tabular}
      \end{minipage}
   \end{table*}}

\newcommand\gitRevision{c82bf8c}
\newcommand\gitAuthorDate{Thu Apr 12 14:44:07 2018 +0200}

\journal{Discrete Applied Mathematics}

\begin{document}

\begin{frontmatter}

\title{Fast Algorithms for Indices of Nested Split Graphs Approximating
Real Complex Networks}
\tnotetext[tn1]{
Manuscript submitted June 22, 2017;
revised March 12, 2018;
accepted March 19, 2018 for publication in
Discrete Applied Mathematics.
}

\author[math]{Irene~Sciriha}
\ead{irene.sciriha-aquilina@um.edu.mt}
\author[cce]{Johann~A.~Briffa\corref{cor1}}
\ead{johann.briffa@um.edu.mt}
\author[math]{Mark~Debono}

\cortext[cor1]{Corresponding author}
\address[math]{Dept.\ of Mathematics,
      University of Malta, Msida MSD~2080, Malta}
\address[cce]{Dept.\ of Commun.\ \& Computer Engineering,
      University of Malta, Msida MSD~2080, Malta}

\begin{abstract}
We present a method based on simulated annealing to obtain a nested split graph
that approximates a real complex graph.
This is used to compute a number of graph indices using very efficient
algorithms that we develop, leveraging the geometrical properties of nested
split graphs.
Practical results are given for six graphs from such diverse areas as social
networks, communication networks, word associations, and molecular chemistry.
We present a critical analysis of the appropriate perturbation schemes that
search the whole space of nested split graphs and the distance functions
that gauge the dissimilarity between two graphs.
\end{abstract}


\end{frontmatter}


\section{Introduction}

The architecture of real-life networks may not be predictable and involves
data which is sometimes difficult to manage.
Algorithms known to date for the computation of many parameters associated
with complex networks are often exponential-time, blocking further development.
To overcome this problem, the strategy implemented here is to use the class
of \emph{nested split graphs} (NSG) since a graph in this class can provide
a model network which is sufficiently close to a given real-world network
and which enables easy computation.
The class of NSGs provides an ideal candidate as these graphs are nice to
work with, are easily stored, and their structured topology lends itself to
polynomial time algorithms for the computation of certain of its invariants.
Where this is not possible, their rich mathematical versatility may enable
other methods to circumvent the problem.

Nested split graphs, also known as \emph{threshold graphs}, form a subclass of split
graphs in which the vertex set is partitioned into a clique (maximal complete
subgraph) and a co-clique, that is an independent subset of vertices (with
no edges between any pair of the independent vertices).
Furthermore, NSGs display a structure in which the clique and the co-clique
are partitioned into \emph{cells} which form an \emph{equitable vertex partition},
that is each vertex of a cell has the same number of neighbours in each of
the cells.
Moreover, the adjacencies among the cells impose a structure on the NSG that
induces a partition of the vertices referred to as the \emph{NSG
partition}.

The questions we ask are:
\begin{inparaenum}[(a)]
\item given a network $G$, is there a NSG $\hat{G}$ that is sufficiently
close to $G$ that can be used instead of $G$ for the purposes of computing
network parameters and spectral properties? and
\item how can computation time be reduced by drawing on the geometric
properties of NSGs?
\end{inparaenum}

To search for the optimum $\hat{G}$, we devise two \emph{distance functions},
the scaled walks and the spectral functions, $f_{SW}$ and $f_{\lambda}$,
respectively, with the objective to measure the rate of convergence of the
original given network to an interim NSG $G'$ picked by the algorithm considered.
The techniques used are Markov chains and simulated annealing that enable
the space of coded NSGs to be searched.
Three perturbation schemes are used for each of the distance functions.
These are the `hamming', `edge', and `move' schemes, all of which will be
explained in detail in Section~\ref{sec:distance}.
The process outputs the best network among all NSGs that can replace the given
network, up to a prescribed tolerance, for the purpose of computing selected
invariants associated with the original network.

The parameters to be computed using the different techniques are derived
mainly from sociology, information theory and physical chemistry.
They are Entropy, the Randi\'c, Wiener, Szeged, Co-PI and Estrada indices,
and also Gutman's Graph Energy and Resolvent Energy.
The networks tested come from communication networks, word associations,
and molecular chemistry.

When using Markov chains and simulated annealing to determine a NSG close
enough to a given $G$, the problem centres on the best distance function and
Markov chain perturbation to use for reasonable computation times and reliable
values of graph invariants of the NSG to which the process converges.
The different techniques are analysed to see which distance functions and
perturbations correspond to optimized output and well behaved algorithms.

The paper will be organised as follows.
In Section~\ref{sec:background} we start by presenting various characterization
of NSGs.
This is followed in Section~\ref{sec:measures} by descriptions of the graph
parameters to be estimated, and the distance functions to be used as a measure
of closeness between the interim NSG $G'$ in the simulated annealing process
and the original graph $G$.
The simulated annealing process is presented in Section~\ref{sec:algorithm},
together with the perturbation schemes used to traverse the whole space of NSGs.
In this section we also develop efficient algorithms to determine the graph
parameters of Section~\ref{sec:measures} by taking advantage of the geometric
properties of NSGs.
Estimates of selected parameters of $G$ are obtained by calculating them
for $\hat{G}$, the limiting NSG.
Practical results are given in Section~\ref{sec:results} for a selection of
six graphs, where we consider the effect of the simulated annealing parameters
on the NSGs obtained, and also compare the indices for the limiting NSGs with
those of the original graphs.
In Section~\ref{sec:closure} pointers for further work from these seminal ideas
are suggested.


\section{Characterizations of a Nested Split Graph}
\label{sec:background}

A complete graph on $n$ vertices is $K_n$ and a graph on $n$ vertices with
no edges is $\overline{K_n}$. The path and the cycle on $n$ vertices are
denoted by $P_n$ and $C_n$, respectively.
A disconnected graph consisting of $r$ copies of a graph $G$ is denoted by $rG$.
The number of edges incident to a vertex $v$ of a graph $G$ on $n$ vertices
is the degree $\rho_v$.
A graph has an isolated vertex if it has a vertex with no edges incident to
it, while a dominating vertex corresponds to a vertex with edges to all the
other vertices of the graph.
A vertex $v$ is a \emph{duplicate} of a vertex $u$ in a graph $G$ if $u$ and
$v$ are not adjacent and they have the same neighbouring vertices in $G$.
Two vertices $u$ and $v$ are \emph{co-duplicates} if they are adjacent and
have the same neighbours.

Different characterizations of NSGs which we shall use in computing the
different NSG parameters are now presented.

\begin{theorem}
\label{ThmNSG1}
The following statements are equivalent for a graph $G$
\cite{sciriha2012spectrum, mahadev1995threshold}:
\begin{enumerate}[(i)]
\item $G$ is a NSG;
\item $G$ is $P_4$, $C_4$ and $2K_2$ free;
\item \label{ThmNSG1:3} $G$ is constructed from $K_1$ by successive additions
of an isolated or dominating vertex.
\end{enumerate}
\end{theorem}

From Theorem~\ref{ThmNSG1}(\ref{ThmNSG1:3}), the following result gives the
construction of a connected NSG from a binary code.

\begin{proposition}
\label{PropCode}
A connected NSG on $n$ vertices is uniquely coded as a binary string of $(n-2)$
bits.
The $r$th bit is 0 if the $(r+1)$th vertex added is an isolated vertex
and it is 1 if it is a dominating vertex.
This string, which is referred to as the \emph{minimum representation} of the
NSG, assumes that the first vertex in the construction is an isolated vertex
and that the last vertex is a dominating vertex.
\end{proposition}

Observe that $(n-2)$ bits are sufficient because in the canonical construction
the first vertex is always isolated and the last vertex has to be dominating.
It also follows from Theorem~\ref{ThmNSG1}(\ref{ThmNSG1:3}) that each
intermediate subgraph in the construction of the NSG $G$ is itself a NSG.
If the first bit of the encoding string (reading from the left) is 0, then
the intermediate NSG $G_2$ on two vertices is $2K_1$, whereas if it is 1,
then $G_2$ is $K_2$.

\begin{definition}
The \emph{creation sequence} for a connected NSG on $n$ vertices is the
$n$-tuple $\mathbf{c}=(c_1,c_2,\ldots,c_n)$ with the first element $c_1=0$,
the last element $c_n=1$, and intermediate elements equal to the bits of
the minimum representation.
\end{definition}

It is convenient to make use of a more concise form of the creation sequence by
encoding repetitions of successive bits.

\begin{definition}
The \emph{compact creation sequence} for a connected NSG on $n$ vertices
is the $r$-tuple $\mathbf{a} = (a_1,a_2,\ldots,a_r)$, where $\sum a_i=n$,
$r$ is even, and for all $i$, $a_i \geq 1$ is the length of the $i$th
successive sequence of entries in the creation sequence of the same value
(i.e.\ with each entry equal to 0 or each entry equal to 1).
\end{definition}

\begin{figure}[tb]
   \centering
   \includegraphics[scale=1]{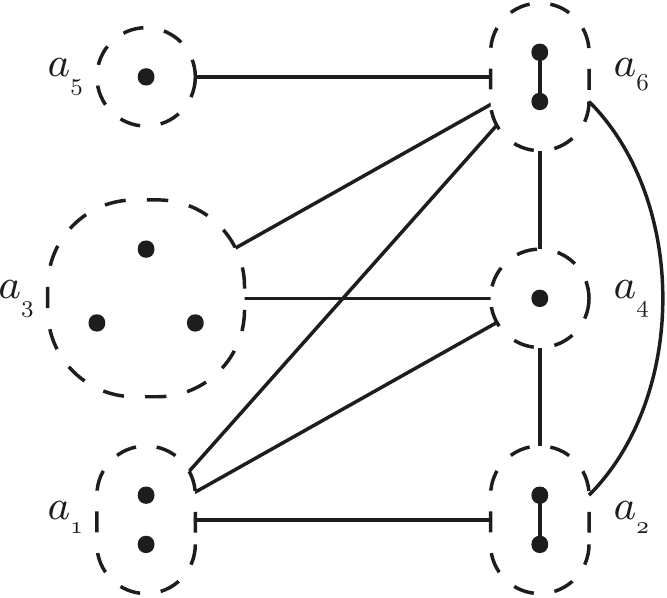}
   \caption{The NSG with compact creation sequence $\mathbf{a}=(2,2,3,1,1,2)$.
   An edge between two cells of the vertex partition indicates that each vertex
   of one cell is adjacent to each vertex of the other cell.}
   \label{fig:nsg-223112}
\end{figure}

For example, the minimum representation 011000101 of length 9, corresponds
to the NSG shown in Figure~\ref{fig:nsg-223112} on 11 vertices, with vertex
partition $\mathbf{a}=(2,2,3,1,1,2)$.

Recall that a clique or a co-clique in the NSG vertex partition is referred to as
a \emph{cell}.
Note that the compact creation sequence is a direct representation of the
NSG partition with the smallest even number $r$ of cells.
Therefore, the coefficients of $\mathbf{a}$ with even subscripts
give the size of the cliques, while those with odd subscripts give the size
of the co-cliques in the NSG.
By considering the construction of the $n$-vertex NSG from the compact
creation sequence $\mathbf{a}=(a_1,a_2,\ldots,a_r)$, the next result follows.
\begin{proposition}
If to a NSG $G$ on $n$ vertices, a duplicate to a vertex in the co-clique or
a co-duplicate to a vertex of the clique is added, then a NSG $G'$ on $n+1$
vertices is obtained with the same number of cells.
\end{proposition}

The adjacency matrix $\mathbf{A}=(A_{i,j})$ of a NSG can be obtained directly
from the creation sequence $\mathbf{c}$ as
\begin{equation}
\mathbf{A} =
   \begin{pmatrix}
   0      & c_2    & c_3    & \cdots & c_n \\
   c_2    & 0      & c_3    & \cdots & c_n \\
   c_3    & c_3    & 0      & \cdots & c_n \\
   \vdots & \vdots & \vdots & \ddots & \vdots \\
   c_n    & c_n    & c_n    & \cdots & 0
   \end{pmatrix}
\end{equation}
Observe that there is no computation involved in this process.


\section{Measures}
\label{sec:measures}

Given a graph $G = (V,E)$, we would like to find a nested split graph (NSG)
$\hat{G}$ that is sufficiently close to $G$.
Closeness in this sense implies that $\hat{G}$ has similar properties to $G$
where it matters within a given context.
Such properties are usually defined in terms of one or more indices of the
graph.
We review the indices on which the algorithms are based.

\subsection{Indices of a graph}
\label{sec:indices}


We consider first indices based on graph topology.
The simplest of these is a function of the number of edges, $|E|$.
Specifically, for an undirected graph $G=(V,E)$ where all edges have
unit weight, the \emph{entropy} \cite{cardinal2004minimum, Dehmer201157,
simonyi1995graph, das2017some} is given by
\begin{equation}
H(G) = \log_2 |E|
\end{equation}


Increasing in complexity, we consider indices that are a function of the
vertex degrees.
The \emph{Randić} index \cite{DVORAK2011434} of a graph $G=(V,E)$ is
given by
\begin{equation}
R(G) = \sum_{uv \in E} \frac{1}{\sqrt{\rho_u \rho_v}}
\end{equation}
where $\rho_u$ and $\rho_v$ are the degrees of vertices $u$ and $v$
respectively.


A further increase in complexity is necessary for indices that are a function
of the shortest distance between vertex pairs.
The simplest of these is the \emph{Wiener} index \cite{Pisanski1991Wiener,
Pisanski1988Wiener, AMC795} of a graph $G=(V,E)$ given by
\begin{equation}
W(G) = \frac{1}{2} \sum_{u,v \in V} d(u,v)
\end{equation}
where $d(u,v)$ is the shortest distance between vertices $u$ and $v$.
%
A related metric is the \emph{Szeged} index \cite{Gutman1995SzegedWiener,
BONAMY2017202} of a graph $G=(V,E)$ given by
\begin{equation}
Sz(G) = \sum_{uv \in E} \nu(u,v) \cdot \nu(v,u)
\end{equation}
where $\nu(u,v) = \left| \{w \mid w \in V, d(u,w) < d(v,w)\} \right|$ is
the number of vertices that are closer to $u$ than to $v$.
%
A variation on the Szeged index, the \emph{Co-PI} index \cite{YuFeng2016CoPi}
of a graph $G=(V,E)$ is given by
\begin{equation}
CoPI(G) = \sum_{uv \in E} \left| \nu(u,v) - \nu(v,u) \right|
\text{.}
\end{equation}


Next we consider indices based on the graph's spectral properties.
The \emph{Estrada} index \cite{delapena2007, li2017new} of a graph $G=(V,E)$
is defined by
\begin{equation}
EE(G) = \sum_{j=1}^{n} e^{\lambda_j}
\end{equation}
where $n=|V|$ and $\lambda_j$, $j=1, \ldots, n$ are the eigenvalues of the
graph's adjacency matrix $\mathbf{A}$.
%
A related metric is Gutman's graph energy \cite{Gutman2012Energy,
BoZhouEnergy2004, das2017energy, vaidya2017some}, obtained as
\begin{equation}
GE(G) = \sum_{j=1}^{n} |\lambda_j|
\end{equation}
which is the sum of absolute values of the eigenvalues of the graph's
adjacency matrix.
%
Finally, the resolvent energy \cite{gutman2016resolvent,
GutmanEnergiesResolvent2016} is also obtained from the eigenvalues of the
graph's adjacency matrix, and is defined as
\begin{equation}
RE(G) = \sum_{j=1}^{n} \frac{1}{n-\lambda_j}
\end{equation}

\subsection{Measures of distance between graphs}
\label{sec:distance}

The problem of finding a NSG $\hat{G}$ that is sufficiently close to a given
graph $G$ can be seen as a combinatorial optimization problem, where the
objective function to be minimized is a measure of the distance between the
two graphs.
This raises the inevitable question of what objective function should be used.
One approach is to choose a graph parameter and use either the absolute difference
or the Euclidean distance between the parameters of $G$ and $\hat{G}$.
An alternative is to choose a vector function that represents the
properties of interest for a graph, and use say the Euclidean distance between the
vectors representing $G$ and $\hat{G}$ as a measure of dissimilarity.
We propose two vector functions, representing the graph's walk and
spectral properties.

The first vector function $\mathbf{W}$ is based on the number of walks
from each vertex, and represents the reachability of vertices from each
other.
Specifically, for a graph $G=(V,E)$ with adjacency matrix $\mathbf{A}$
we define the scaled walk matrix by
\begin{equation}
\mathbf{W} = \left( W_0 \| W_1 \| \cdots \| W_{n-1} \right)
\end{equation}
which is the concatenation of scaled walk vectors $W_i$ for walks of length
$i$, given by
\begin{equation}
W_i = \frac{ \mathbf{A}^i\mathbf{j} }{ \Delta^i}
\end{equation}
where $\Delta = \max_{v \in V} \rho_v$ is the maximum vertex degree of $G$ and
$\mathbf{j}$ is the all-one vector of length $n = |V|$.

The second vector function $\mathbf{\Lambda}$ represents the graph's spectral
properties.
Specifically, for a graph $G=(V,E)$ with adjacency matrix $\mathbf{A}$
we use the vector of eigenvalues \cite{SciConstrNullOne98, SciCoefx97,
SciCHznSingGr07} defined by
\begin{equation}
\mathbf{\Lambda} = \left( \lambda_1, \lambda_2, \ldots, \lambda_n \right)
\end{equation}
where $n=|V|$ and $\lambda_j$, $j=1, \ldots, n$ are the eigenvalues of
$\mathbf{A}$ indexed in non-increasing order.


\section{Algorithm}
\label{sec:algorithm}

\subsection{Simulated annealing}
\label{sec:simulatedannealing}

Simulated annealing is a probabilistic global optimization algorithm that
mimics the physical process of thermal annealing, by which crystals are
slowly cooled to reach a state with the lowest free energy.
In analogy, the algorithm seeks the lowest energy state from a large but
finite state space, where the energy is a deterministic function of the state.

The algorithm traverses a Markov chain, where the probability of accepting
a state change depends on the difference in energy $\Delta_E$ between the
current and candidate states, and on a time-varying parameter $T$ called
the temperature.
Specifically, when the candidate state has a lower energy than that of
the current state, then $\Delta_E < 0$ and the acceptance probability is
$1$, while when the candidate state has an energy equal to or higher
than that of the current state, this is accepted with probability
$e^{-\frac{\Delta_E}{T}}$.
The initial temperature needs to be large enough so that a transition to a
higher-energy state is allowed with high probability.
As the temperature is reduced, such transitions are accepted with lower
probability.
The final temperature is chosen so that only transitions that reduce the
energy are allowed with any significant probability.
Typically the annealing schedule follows a geometric decrease from the
initial to final temperature.

\subsection{Perturbation schemes}
\label{sec:perturbations}

Recall that for a NSG, the creation sequence $\mathbf{c}=(c_1,c_2,\ldots, c_n)$
must satisfy the constraints $c_1=0$ and $c_n=1$.
Therefore, we define the state variable that uniquely corresponds to this
NSG as the tuple $\sigma=(c_2,\ldots,c_{n-1})$.

The transition probabilities of the state space define a neighbourhood,
where ideally the neighbours of a state have an energy similar to that of
the initial state.
The neighbourhood must be defined in such a way that, while traversing
the Markov chain, all states are reachable, following a sufficient number
of transitions.
This ensures that the whole state space can be explored.
The neighbourhood can be defined as a perturbation function on the
state variable.
We propose three different perturbation schemes: `hamming', `edge', and
`move', that define neighbourhoods within the state space.

With the `hamming' scheme, the neighbourhood is defined as the set of those nested
split graphs with a valid creation sequence at a Hamming distance of one
from the current state's creation sequence.
The neighbouring states are those corresponding to creation sequences
$\mathbf{c}'=(c'_1,c'_2,\ldots,c'_n)$, where
\begin{align}
c'_i &=
   \begin{cases}
   \overline{c_i} & \text{if~} i=j \\
   c_i & \text{otherwise,} \\
   \end{cases}\\
\overline{c} &=
   \begin{cases}
   0 & \text{if~} c=1 \\
   1 & \text{if~} c=0 \\
   \end{cases}
\end{align}
and $j \in \{2,\ldots,n-1\}$ identifies the specific neighbour.
Observe that this definition creates a path between all possible states,
with a minimum distance of at most $n-2$ steps.
In our implementation, the neighbourhood is uniformly sampled.

Observe also that with the `hamming' scheme not all neighbouring graphs are
equally similar to the current state.
For example, consider the current state corresponding to
$\mathbf{c}=(0,1,1,1,1)$, equivalent to $K_5$.
One possible neighbour corresponds to $\mathbf{c}'=(0,0,1,1,1)$, which
has one less edge than $\mathbf{c}$, while another corresponds to
$\mathbf{c}'=(0,1,0,1,1)$, which has two fewer edges than $\mathbf{c}$.
In order to minimise the differences between neighbours, we define the `edge'
scheme as the perturbation whose neighbourhoods form a subset of the `hamming'
neighbourhood where the changed coefficient is the first or last variable
coefficient or it is different from either of its adjacent coefficients.
Thus
\begin{equation*}
j \in \{2,n-1\} \cup
  \{i : (c_i \neq c_{i-1}) \text{~or~} (c_i \neq c_{i+1}) \text{~for~} 2<i<n-1 \}
\text{.}
\end{equation*}
The latter condition allows the movement of a vertex from a cell to another
cell with adjacent subscript, that is from a clique $2k$ to a co-clique $2k-1$
or $2k+1$ and from a co-clique $2k-1$ to a clique $2k$ or $2k-2$.
This includes cases where, as a result of the change, the number of vertices
in the initial cell is reduced to zero, and forces the consequent merging
of both cells that are adjacent in index to the vanishing one.
This condition provides paths to all states with equal or smaller number
of cells than the current state.
The former condition allows the inversion of coefficients $c_2$ and
$c_{n-1}$, and provides paths to states with an increased number of cells.

Finally, we define the `move' scheme, where neighbouring states are obtained
from the current one by moving one of the vertices in a source cell $j$
to a different target cell $k$ where $k$ is at most two different from $j$.
Formally, for a current state defined by the compact creation sequence
$\mathbf{a} = (a_1,a_2,\ldots,a_r)$, neighbouring states are
those corresponding to compact creation sequences
$\mathbf{a}' = (a'_1,a'_2,\ldots,a'_r)$, where
\begin{equation}
a'_i =
\begin{cases}
a_i-1 & \text{if~} i=j \\
a_i+1 & \text{if~} i=k \\
a_i   & \text{otherwise} \\
\end{cases}
\end{equation}
and $1 \leq j \leq r$, $j \in \mathbb{N}$ is the cell a vertex is moved from
while $1 \leq k \leq r$, $k \in \{ j-2, j-1, j+1, j+2 \}$ is the target cell.
This condition provides paths to all states with equal or smaller number
of cells than the current state.

The neighbourhood also includes states with $r+2$ cells, obtained by moving
two vertices from cell $j$ to form two new cells of size 1.
These states correspond to compact creation sequences
$\mathbf{a}'' = (1,1,a''_1,a''_2,\ldots,a''_r)$ and
$\mathbf{a}''' = (a''_1,a''_2,\ldots,a''_{r-1},a''_r,1,1)$, where
\begin{equation}
a''_i =
\begin{cases}
a_i-2 & \text{if~} i=j \\
a_i   & \text{otherwise} \\
\end{cases}
\end{equation}
and $a_j \geq 2$, $j \in \{1,2\}$ in the case of $\mathbf{a}''$,
while $j \in \{r-1,r\}$ in the case of $\mathbf{a}'''$.
Note that the above definition includes degenerate compact creation sequences,
where the size of one cell becomes zero.
In these cases, both cells that are adjacent in index to the vanishing
one will merge.

\subsection{Fast computation of indices on NSGs}
\label{sec:nsgindices}

The structure of NSGs lends itself to less complex algorithms for computing
a number of graph indices.
We consider here the indices introduced in Section~\ref{sec:indices},
describing algorithms to compute these for NSGs, and comparing the complexity
of these algorithms with those for general graphs.


We start with the computation of entropy.
This requires the computation of the number of edges in a graph, which for
a general simple undirected graph can be computed as:
\begin{equation}
m = \sum_{i=2}^{n} \sum_{j=1}^{i-1} A_{i,j}
\end{equation}
where $A_{i,j}$ is the element at row $i$, column $j$ of $\mathbf{A}$.
This is a summation over $\frac{n^2-n}{2}$ elements, for a complexity $O(n^2)$.
For a NSG with compact creation sequence $\mathbf{a} = (a_1,a_2,\ldots,a_r)$,
the number of edges can be computed as a summation of two terms using the elements of
$\mathbf{a}$:
\begin{equation}
m = \frac{\kappa_1(\kappa_1-1)}{2}
  + \sum_{k=1}^{r/2} \left( a_{2k-1} \kappa_k \right)
\end{equation}
where $\kappa_k = \sum_{j=k}^{r/2} a_{2j}$.
The first term gives the number of edges in the subgraph containing all
cliques, while the second term considers the edges between each co-clique
and all the cliques it is connected to.
This requires the computation of the cumulative sum of even-indexed elements
followed by a summation over the odd-indexed elements of $\mathbf{a}$,
for a complexity $O(r)$.
The computation of entropy follows directly, with constant complexity.


Consider next the Randić index, which is a function of the degrees of the
vertices at either end of an edge.
Observe that all vertices within a cell have the same degree.
For co-clique $a_{2k-1}$ on level $k \in \{1,\ldots,r/2\}$ the degree is equal to
the sum of vertices in cliques at the same level or higher:
\begin{equation}
\rho_{2k-1} = \sum_{j=k}^{r/2} a_{2j}
\text{.}
\end{equation}
Similarly, for clique $a_{2k}$, the degree is equal to the sum of vertices
in co-cliques at the same level or lower plus the number of other vertices
in cliques:
\begin{equation}
\rho_{2k} = \sum_{j=1}^{k} a_{2j-1} + \sum_{j=1}^{r/2} a_{2j} - 1
\text{.}
\end{equation}
Each vertex degree requires computation of complexity $O(r)$, for a total
complexity $O(r^2)$ to compute all the vertex degrees.
The Randić index can then be obtained by summing over the edges between
cells and the edges within cliques, taking into account the multiplicity in
each case:
\begin{equation}
\label{eq:randic_direct}
\begin{split}
R(G) &=
   \sum_{k=1}^{r/2}
      \sum_{j=k}^{r/2}
         \frac{a_{2k-1} a_{2j}}{\sqrt{\rho_{2k-1} \rho_{2j}}}
 + \sum_{k=1}^{r/2-1}
      \sum_{j=k+1}^{r/2}
         \frac{a_{2k} a_{2j}}{\sqrt{\rho_{2k} \rho_{2j}}}
 + \sum_{k=1}^{r/2}
      \frac{a_{2k} (a_{2k}-1)}{2 \rho_{2k}}
\text{.}
\end{split}
\end{equation}
Respectively, the terms count over the edges between co-cliques and cliques,
edges between distinct cliques, and edges within cliques.
Computationally, this can also be improved to
\begin{equation}
R(G) =
   \sum_{k=1}^{r/2}
      \alpha_{2k-1}
      \beta_k
 + \sum_{k=1}^{r/2-1}
      \alpha_{2k}
      \beta_{k+1}
 + \sum_{k=1}^{r/2}
      \frac{a_{2k} (a_{2k}-1)}{2 \rho_{2k}}
\end{equation}
where $\alpha_k = \frac{a_k}{\sqrt{\rho_k}}$ and
$\beta_k = \sum_{j=k}^{r/2} \alpha_{2j}$.
This involves four computations of complexity $O(r)$, reducing the
complexity from $O(r^2)$ when \eqref{eq:randic_direct} is used.


The Wiener index introduces a further increase in complexity, requiring the
shortest distance between every vertex pair.
For a general undirected graph, this can be obtained with a breadth-first
search starting from every vertex, for an overall complexity $O(n^3)$.
For a NSG we can observe that:
\begin{inparaenum}[(a)]
\item the distance between two vertices in cliques is always equal to one,
\item the distance between two vertices in co-cliques is always equal to two
(they are always connected to at least one common clique vertex),
\item the distance between a vertex in a co-clique and a vertex in a clique
at the same level or higher is always equal to one, and
\item the distance between a vertex in a co-clique and a vertex in a clique
at a lower level is always equal to two.
\end{inparaenum}
Now the Wiener index can be computed as the sum of the shortest distance
between each distinct pair of vertices, which leads to
\begin{equation}
\begin{split}
W(G) &=
   \sum_{k=1}^{r/2}
      \frac{ a_{2k} (a_{2k}-1) }{2}
   + \sum_{k=1}^{r/2-1}
      a_{2k}
      \sum_{j=k+1}^{r/2}
         a_{2j}
   + \sum_{k=1}^{r/2}
      a_{2k-1} (a_{2k-1}-1) \\
   &\quad + \sum_{k=1}^{r/2-1}
      2 a_{2k-1}
      \sum_{j=k+1}^{r/2}
         a_{2j-1}
   + \sum_{k=1}^{r/2}
      a_{2k-1}
      \sum_{j=k}^{r/2}
         a_{2j}
   + \sum_{k=1}^{r/2-1}
      2 a_{2k}
      \sum_{j=k+1}^{r/2}
         a_{2j-1}
\text{.}
\end{split}
\end{equation}
As before, this can be simplified to
\begin{equation}
\begin{split}
W(G) &=
   \sum_{k=1}^{r/2} \left(
      a_{2k-1} (a_{2k-1}-1 + \delta_k)
      + \frac{ a_{2k} (a_{2k}-1) }{2}
      \right)
   + \sum_{k=1}^{r/2-1} \left(
      2 a_{2k-1} \gamma_{k+1}
      + a_{2k} (\delta_{k+1} + 2 \gamma_{k+1})
      \right)
\end{split}
\end{equation}
where $\gamma_k = \sum_{j=k}^{r/2} a_{2j-1}$ and
$\delta_k = \sum_{j=k}^{r/2} a_{2j}$.
Again this reduces the computation to complexity $O(r)$.


The Szeged index first requires the computation of $\nu(u,v)$, the number
of vertices that are closer to $u$ than $v$, for every directly connected
pair of vertices $u,v$ in the graph.
Now for a NSG there are three types of directly connected vertices:
\begin{inparaenum}[(a)]
\item when both in the same clique,
\item when the vertices are in different cliques, and
\item \label{type_c} when one vertex is in a co-clique and the other is in a clique at the
same level or higher.
\end{inparaenum}
Consider these in turn, observing that $\nu(u,v) \geq 1$ as there will always
be at least one vertex ($u$ itself) that is closer to $u$ than $v$.
If $u,v$ are in the same clique, they are co-duplicates; so it follows
that $\nu(u,v)=\nu(v,u)=1$.
Without loss of generality, if $u$ is in clique $2k$ and $v$ is in clique $2j$
where $k<j$ and $k,j \in \{1,\ldots,r/2\}$, then $u$ has the same connections
as $v$ except for co-cliques $a_{2i-1}$, $i \in \{k+1,\ldots,j\}$, where $v$
has a direct connection while $u$ does not.
It follows, in this case, that $\nu(u,v)=1$ and $\nu(v,u)=1 + \sum_{i=k+1}^{j}
a_{2i-1}$.
Finally, if $u$ is in co-clique $2k-1$ and $v$ is in clique $2j$ where $k
\leq j$ and $k,j \in \{1,\ldots,r/2\}$, then we need to consider the distances
from $u,v$ to other vertices $w$ in the graph:
\begin{inparaenum}[(i)]
\item if $w$ is in a co-clique at the level of $v$ or lower (including the
same co-clique as $u$), it is adjacent to $v$ but not to $u$, so that
$d(v,w) < d(u,w)$,
except when $w = u$, in which case $d(u,w) < d(v,w)$
\item if $w$ is in a co-clique at a level above $v$, it is adjacent to neither
$u$ nor $v$, and $d(v,w) = d(u,w) = 2$,
\item if $w$ is in a clique at the level of $u$ or higher (including the
same clique as $v$), it is adjacent to both $u$ and $v$, so that
$d(v,w) = d(u,w) = 1$,
except when $w = v$, in which case $d(v,w) < d(u,w)$,
and
\item if $w$ is in a clique at a level below $u$, it is adjacent to $v$
but not to $u$, so that $d(v,w) < d(u,w)$.
\end{inparaenum}
Combining these for case (\ref{type_c}), it follows that
$\nu(u,v)=1$ and
$\nu(v,u)=\sum_{i=1}^{j} a_{2i-1}$ for $k=1$, while
$\nu(v,u)=\sum_{i=1}^{j} a_{2i-1} + \sum_{i=1}^{k-1} a_{2i}$ for $k>1$.
Finally, to obtain the Szeged index we need to sum over all edges of the graph:
\begin{equation}
\begin{split}
Sz(G) &=
   \sum_{k=1}^{r/2}
      \frac{a_{2k} (a_{2k}-1)}{2}
   + \sum_{k=1}^{r/2-1}
      a_{2k}
      \sum_{j=k+1}^{r/2}
         a_{2j}
         \left( 1 + \sum_{i=k+1}^{j} a_{2i-1} \right) \\
   &\quad +
      a_{1}
      \sum_{j=1}^{r/2}
         a_{2j}
         \sum_{i=1}^{j} a_{2i-1}
   + \sum_{k=2}^{r/2}
      a_{2k-1}
      \sum_{j=k}^{r/2}
         a_{2j}
         \left( \sum_{i=1}^{j} a_{2i-1} + \sum_{i=1}^{k-1} a_{2i} \right)
\text{.}
\end{split}
\end{equation}
For computational purposes, this can be simplified to
\begin{equation}
\begin{split}
Sz(G) &=
   \sum_{k=1}^{r/2}
      \frac{a_{2k} (a_{2k}-1)}{2}
   + \sum_{k=1}^{r/2-1}
      a_{2k}
      \left( \delta_{k+1} (1 - \epsilon_k) + \eta_{k+1} \right)
   + a_{1} \eta_1
   + \sum_{k=2}^{r/2}
      a_{2k-1}
      \left( \eta_{k} + \delta_k \zeta_{k-1} \right)
\end{split}
\end{equation}
where $\epsilon_k = \sum_{j=1}^{k} a_{2j-1}$,
$\zeta_k = \sum_{j=1}^{k} a_{2j}$,
$\eta_k = \sum_{j=k}^{r/2} a_{2j} \epsilon_j$, and
$\delta_k$ is as defined earlier.
This involves a sequence of computations of complexity $O(r)$.


The Co-PI index is also based on $\nu(u,v)$; following earlier observations
for the Szeged index, we can similarly obtain the Co-PI index as a sum over
all edges in the graph:
\begin{equation}
\begin{split}
CoPI(G) &=
   \sum_{k=1}^{r/2-1}
      a_{2k}
      \sum_{j=k+1}^{r/2}
         a_{2j}
         \sum_{i=k+1}^{j} a_{2i-1}
   +  a_{1}
      \sum_{j=1}^{r/2}
         a_{2j}
         \left( \sum_{i=1}^{j} a_{2i-1} - 1 \right)
   + \sum_{k=2}^{r/2}
      a_{2k-1}
      \sum_{j=k}^{r/2}
         a_{2j}
         \left( \sum_{i=1}^{j} a_{2i-1} + \sum_{i=1}^{k-1} a_{2i} - 1\right)
\text{.}
\end{split}
\end{equation}
As for the Szeged index, this can be simplified to
\begin{equation}
\begin{split}
CoPI(G) &=
   \sum_{k=1}^{r/2-1}
      a_{2k}
      \left( \eta_{k+1} - \delta_{k+1} \epsilon_k \right)
   +  a_{1}
      \left( \eta_1 - \delta_1 \right)
   + \sum_{k=2}^{r/2}
      a_{2k-1}
      \left( \delta_k (\zeta_{k-1} - 1) + \eta_{k} \right)
\end{split}
\end{equation}
where $\epsilon_k$, $\zeta_k$, $\eta_k$, and $\delta_k$ are as defined earlier.
Again, this involves a sequence of computations of complexity $O(r)$.


The remaining indices are functions of the eigenvalues of the graph's
adjacency matrix $\mathbf{A}$.
These eigenvalues are referred to as \emph{main} if they have some eigenvector
not orthogonal to the all-one vector $\mathbf{j}$, and \emph{non-main} otherwise
\cite{sciriha2012spectrum}.
Now for a NSG it can be shown that the main eigenvalues of $\mathbf{A}$
can be obtained by computing the eigenvalues of $\mathbf{Q}$ where
$\mathbf{Q}$ is the $r \times r$ matrix whose rows are the distinct rows
of $\mathbf{A}\mathbf{X}$, for $\mathbf{X}$ being an $n \times r$ indicator
matrix satisfying $\mathbf{A}\mathbf{X} = \mathbf{X}\mathbf{Q}$
\cite{sciriha2012spectrum}.
The matrix $\mathbf{Q}$ can be obtained directly from the compact creation
sequence $\mathbf{a}$ as
\begin{equation}
\mathbf{Q} =
   \begin{pmatrix}
   0      & a_2    & 0      & a_4    & \cdots & a_r \\
   a_1    & a_2-1  & 0      & a_4    & \cdots & a_r \\
   0      & 0      & 0      & a_4    & \cdots & a_r \\
   a_1    & a_2    & a_3    & a_4-1  & \cdots & a_r \\
   \vdots & \vdots & \vdots & \vdots & \ddots & \vdots \\
   a_1    & a_2    & a_3    & a_4    & \cdots & a_r-1
   \end{pmatrix}
\text{.}
\end{equation}
Unlike $\mathbf{A}$, $\mathbf{Q}$ is not a symmetric matrix, so the computation
of its eigenvalues is more complex by some constant factor.
However, this is more than made up for by the significant reduction in size from
$n \times n$ to $r \times r$.
This reduces the complexity of computing the eigenvalues from $O(n^3)$ to $O(r^3)$.
In addition to the eigenvalues of $\mathbf{Q}$, $\mathbf{A}$ has two distinct
non-main eigenvalues:
\begin{inparaenum}[(i)]
\item the eigenvalue $-1$ for co-duplicates, with multiplicity
$\sum_{k=1}^{r/2} \left( a_{2k} - 1 \right) = \delta_1 - \frac{r}{2}$, and
\item the eigenvalue zero for duplicates, with multiplicity
$\sum_{k=1}^{r/2} \left( a_{2k-1} - 1 \right) = \gamma_1 - \frac{r}{2}$.
\end{inparaenum}

We can therefore obtain the Estrada index as a summation
\begin{equation}
EE(G) =
   \left( \delta_1 - \frac{r}{2} \right) e^{-1}
   + \left( \gamma_1 - \frac{r}{2} \right)
   + \sum_{j=1}^{r} e^{\hat\lambda_j}
\end{equation}
where $\hat\lambda_j$, $j=1,\ldots,r$, are the eigenvalues of $\mathbf{Q}$.
%
Similarly, Gutman's graph energy is obtained as
\begin{equation}
GE(G) =
   \delta_1 - \frac{r}{2}
   + \sum_{j=1}^{r} |\hat\lambda_j|
\end{equation}
%
and the resolvent energy is obtained as
\begin{equation}
RE(G) =
   \left( \delta_1 - \frac{r}{2} \right) \frac{1}{n+1}
   + \left( \gamma_1 - \frac{r}{2} \right) \frac{1}{n}
   + \sum_{j=1}^{r} \frac{1}{n-\hat\lambda_j}
\text{.}
\end{equation}


\section{Results}
\label{sec:results}

\subsection{Graphs used}
\label{sec:graphs}

\begin{figure}[!tp]
   \centering
   \includegraphics[scale=0.5]{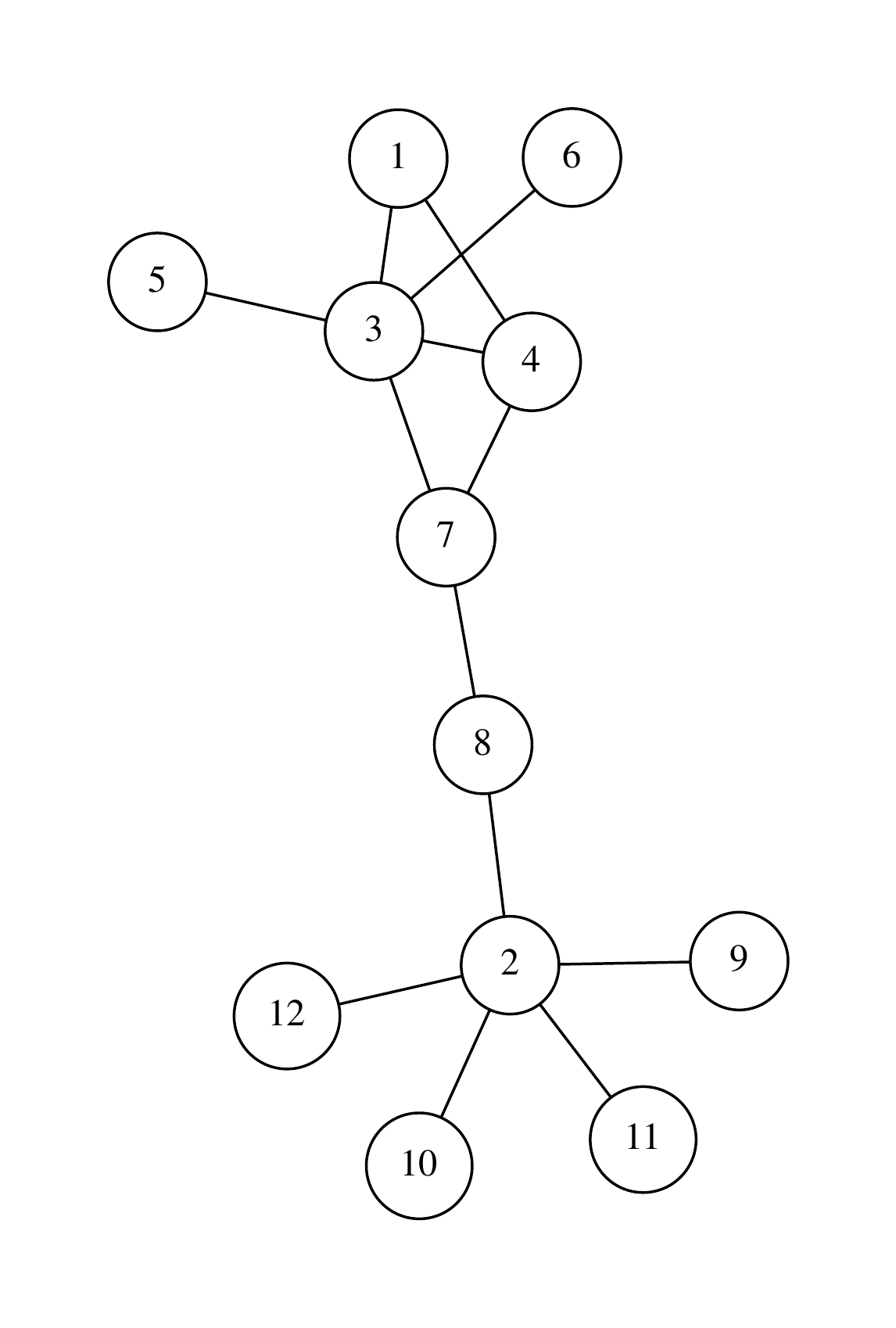}
   \caption{Topology for graph G12}
   \label{fig:G12-topology}
\end{figure}

\begin{figure}[!tp]
   \centering
   \includegraphics[scale=0.5]{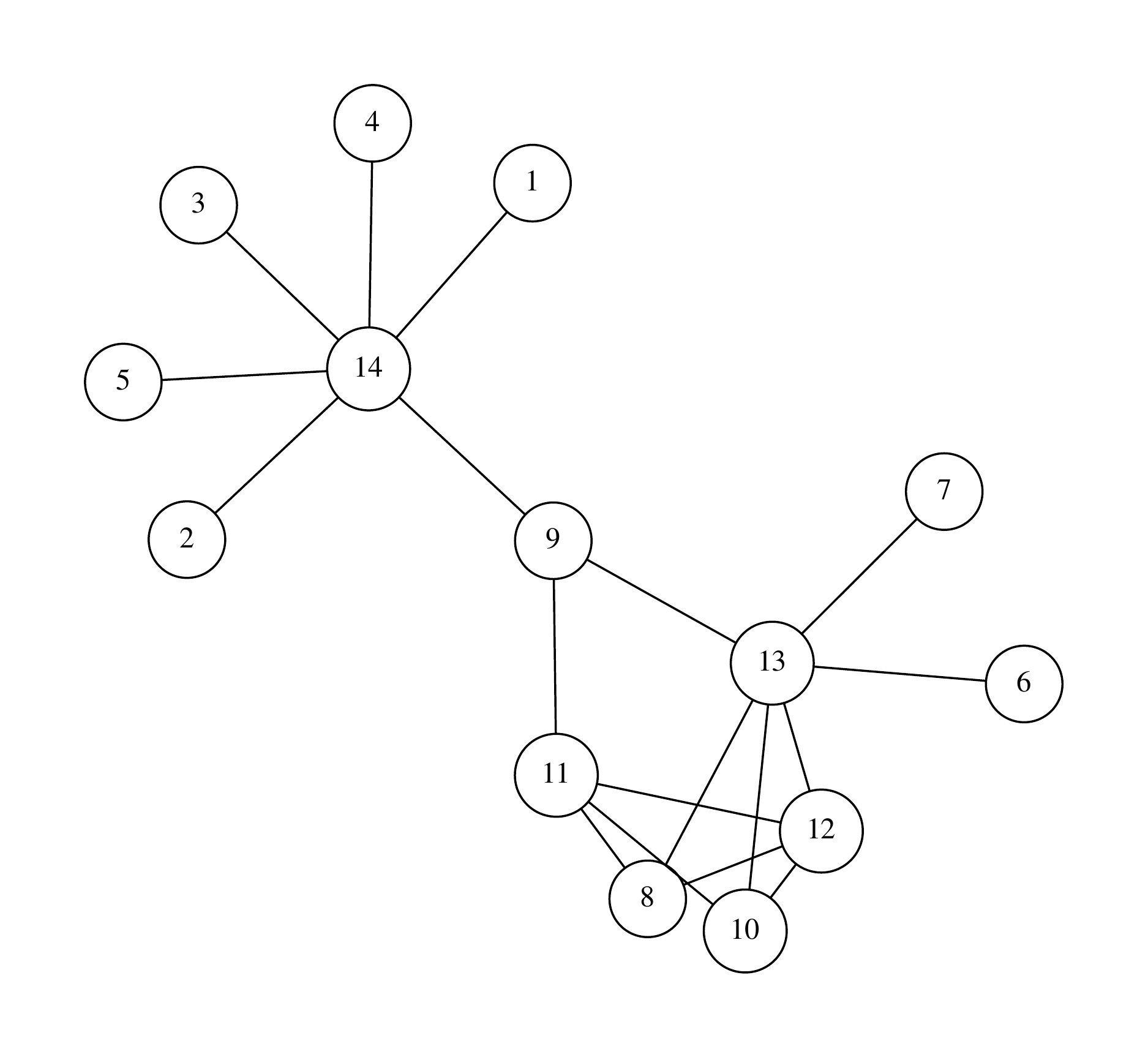}
   \caption{Topology for graph G14}
   \label{fig:G14-topology}
\end{figure}

\begin{figure}[!tp]
   \centering
   \includegraphics[width=0.5\columnwidth]{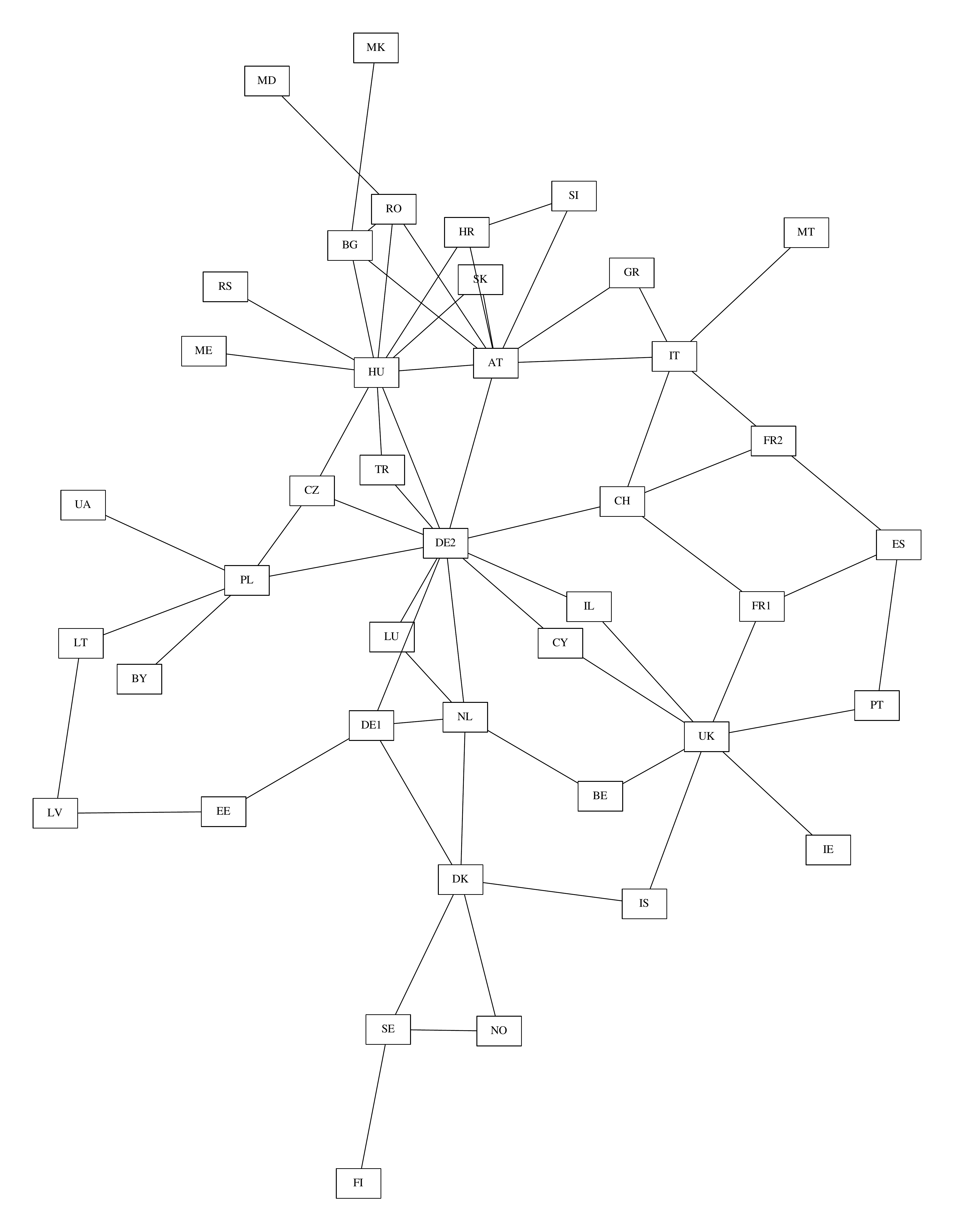}
   \caption{Topology for G\'EANT graph.}
   \label{fig:geant-topology}
\end{figure}

\begin{figure}[!tp]
   \centering
   \includegraphics[width=0.5\columnwidth]{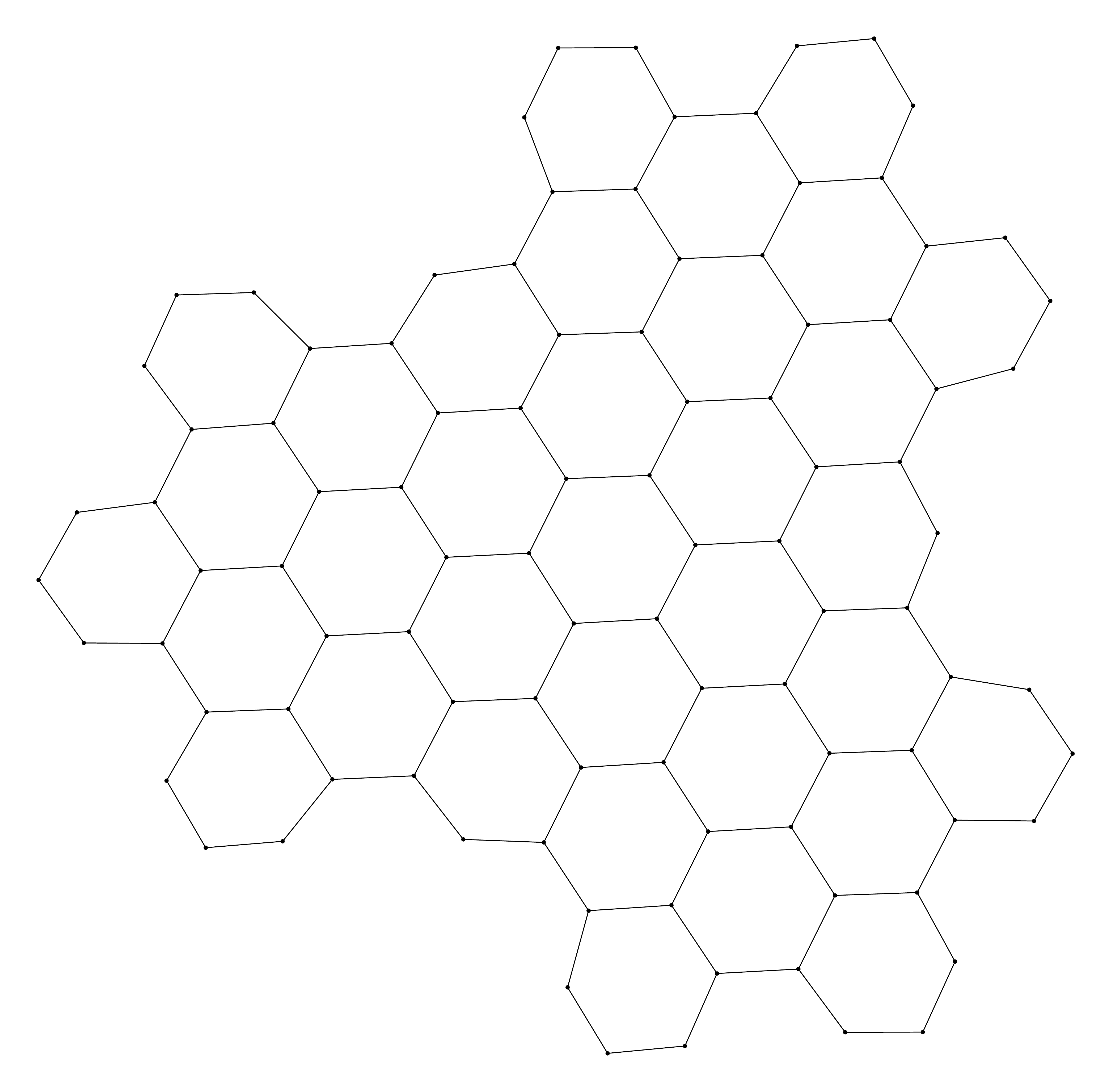}
   \caption{Topology for benzenoid graph.}
   \label{fig:benzenoid-topology}
\end{figure}

Results in this section are for a number of simple connected graphs.
We start with two small graphs taken from social communities, the results
for which can be more easily verified.
These are named G12 and G14, respectively with 12 and 14 vertices, 13 and
18 edges, and topologies as shown in Figures~\ref{fig:G12-topology} and
\ref{fig:G14-topology} respectively.
Next we consider the graph `geant', representing the G\'EANT network topology
as of January 2017 \cite{geant}.
This has 41 vertices and 64 edges, with topology as shown in
Figure~\ref{fig:geant-topology}.
We also consider the graphs `abstract' and `honey', representing word
associations in text paragraphs.
These have 59 and 77 vertices, 605 and 687 edges, respectively.
Finally we consider the graph `benzenoid', representing the carbon bonds
of the large benzenoid known as the \emph{supernaphthalene} made by Klaus
M\"ullen \cite{fowlerprivate}.
This has 96 vertices and 129 edges, with topology as shown in
Figure~\ref{fig:benzenoid-topology}.

\subsection{Simulated annealing}

For each of the graphs described in Section~\ref{sec:graphs}, we use simulated
annealing to find a similar nested split graph, by minimising one of the
distance functions.
Two graph distance functions are used:
\begin{inparaenum}
\item the walk matrix distance, which is the Euclidean distance between the
two graphs' scaled walk matrices, and
\item the spectral distance, which is the Euclidean distance between the two
graphs' eigenvalues.
\end{inparaenum}
For each case, we also use three different perturbation schemes as defined
in Section~\ref{sec:perturbations}.
Throughout we use an annealing schedule where the temperature is varied
logarithmically from $10^2$ to $10^{-7}$ over one million steps.

The final output of the simulated annealing algorithm is tabulated in
Table~\ref{tab:annealing} for graphs G12, G14, `geant', `abstract', `honey',
and `benzenoid'.
\inserttabd{tab:annealing}{ccccl}
{Nested split graph obtained with different perturbation schemes and distance functions,
for graphs G12, G14, `geant', `abstract', `honey', and `benzenoid'.
The closest nested split graph for a given distance function is marked (*).}{
\emph{Graph} & \emph{Function} & \emph{Perturbation} & \emph{Distance} & \emph{$\mathbf{a}$} \\
\hline
\multirow{2}{*}{\rotatebox[origin=c]{90}{G12}}
& walk     & all & 0.3119 & $[1, 2, 1, 1, 5, 2]$ \\
\cline{2-5}
& spectral & all & 1.930  & $[1, 1, 1, 1, 7, 1]$ \\
\hline
\multirow{3}{*}{\rotatebox[origin=c]{90}{G14}}
& walk     & hamming & 0.4285* & $[1, 2, 2, 1, 1, 1, 4, 2]$ \\
& walk     & move /
             edge    & 0.4290  & $[1, 2, 1, 1, 2, 1, 4, 2]$ \\
\cline{2-5}
& spectral & all     & 1.709   & $[1, 2, 2, 1, 7, 1]$ \\
\hline
\multirow{5}{*}{\rotatebox[origin=c]{90}{geant}}
& walk     & hamming & 0.5128* & $[2, 1, 1, 1, 1, 1, 1, 1, 2, 1, 1, 1,
                                   1, 1, 2, 1, 6, 1, 7, 1, 4, 1, 1, 1]$ \\
& walk     & move    & 0.5297  & $[3, 2, 1, 2, 4, 2, 2, 2, 10, 2, 9, 2]$ \\
& walk     & edge    & 0.5305  & $[3, 4, 4, 2, 4, 2, 5, 1, 11, 2, 2, 1]$ \\
\cline{2-5}
& spectral & hamming /
             move    & 7.590*  & $[1, 2, 1, 1, 1, 1, 2, 1, 4, 1, 25, 1]$ \\
& spectral & edge    & 7.608   & $[1, 2, 1, 1, 2, 1, 4, 1, 27, 1]$ \\
\hline
\multirow{4}{*}{\rotatebox[origin=c]{90}{abstract}}
& walk     & hamming & 1.378  & $[1, 1, 2, 1, 1, 2, 4, 1, 1, 1, 1, 2, 1,
                                  1, 10, 2, 5, 1, 8, 1, 1, 1, 3, 1, 4, 2]$ \\
& walk     & move    & 1.378* & $[1, 1, 1, 2, 2, 2, 2, 1, 1, 1, 2, 2, 2,
                                  1, 11, 3, 12, 3, 7, 2]$ \\
& walk     & edge    & 1.383  & $[2, 4, 6, 6, 13, 3, 13, 3, 7, 2]$ \\
\cline{2-5}
& spectral & all     & 16.83  & $[1, 13, 5, 4, 9, 3, 21, 3]$ \\
\hline
\multirow{7}{*}{\rotatebox[origin=c]{90}{honey}}
& walk     & hamming & 1.378* & $[1, 1, 2, 1, 2, 1, 3, 3, 2, 1, 2, 1, 1, 2, 2, 1,$ \\
&&&&                             $2, 1, 1, 1, 1, 4, 2, 3, 11, 2, 2, 1, 2, 1, 16, 1]$ \\
& walk     & move    & 1.378  & $[2, 1, 2, 1, 2, 2, 3, 3, 3, 2, 3, 2, 4, 9,
                                  15, 4, 18, 1]$ \\
& walk     & edge    & 1.379  & $[1, 1, 3, 2, 3, 3, 6, 5, 6, 9, 15, 4, 18, 1]$ \\
\cline{2-5}
& spectral & hamming & 20.52* & $[1, 15, 5, 2, 2, 1, 5, 2, 11, 2, 28, 1, 1, 1]$ \\
& spectral & move    & 20.53  & $[1, 16, 7, 3, 12, 2, 8, 1, 23, 1, 2, 1]$ \\
& spectral & edge    & 20.62  & $[1, 5, 1, 10, 7, 4, 14, 2, 31, 2]$ \\
\hline
\multirow{5}{*}{\rotatebox[origin=c]{90}{benzenoid}}
& walk     & hamming & 11.31  & $[2, 4, 3, 2, 2, 2, 1, 1, 2, 2, 7, 68]$ \\
& walk     & move    & 11.30* & $[1, 1, 1, 2, 4, 2, 3, 4, 9, 69]$ \\
& walk     & edge    & 11.30  & $[1, 2, 4, 3, 4, 4, 9, 69]$ \\
\cline{2-5}
& spectral & hamming /
             move    & 17.34* & $[1, 2, 1, 1, 2, 1, 5, 1, 81, 1]$ \\
& spectral & edge    & 17.36  & $[1, 2, 1, 1, 1, 1, 3, 1, 84, 1]$ \\
}
Comparing the behaviour of the graph distance functions as optimization
targets, we note that the spectral distance is more stable, often converging
to the same graph for different perturbations.
The most significant exception is for the graph `honey'; perhaps this is
because `honey' is not connected while all the other graphs are connected.
On the other hand, the walk matrix distance converges to different graphs,
except for the smallest cases (G12 and G14).

When we consider the effect of the perturbation scheme on the simulated annealing
algorithm, we can draw the following conclusions.
The `hamming' scheme consistently performs best, resulting in the lowest energy case every
time there is a difference between methods, with the exception of `benzenoid'
where the convergent distance for the three methods is very similar.
This comes at a cost: since the `hamming' neighbourhood contains a greater number
of NSGs that are significantly different from the current state, the acceptance
rate during annealing falls more rapidly.
In turn, this requires a slower reduction in temperature, as shown later in the
annealing timeline.
As expected, `move' is consistently better than `edge', as it has a
substantially wider neighbourhood, without including substantially different
NSGs.
We can also see that `edge' favours more compact creation sequences, while
`hamming' results in longer ones.
This can be explained by observing that neighbours in `edge' generally have
the same number of cliques and co-cliques.
An increase only happens with an insertion at either end of the creation
sequence, while a decrease only happens with a move from a clique or co-clique
with one vertex.
As expected, the performance of 'move' is somewhere in between.

\begin{figure}[!tp]
   \centering
   \begin{subfigure}[b]{0.49\linewidth}
      \includegraphics[width=\columnwidth]{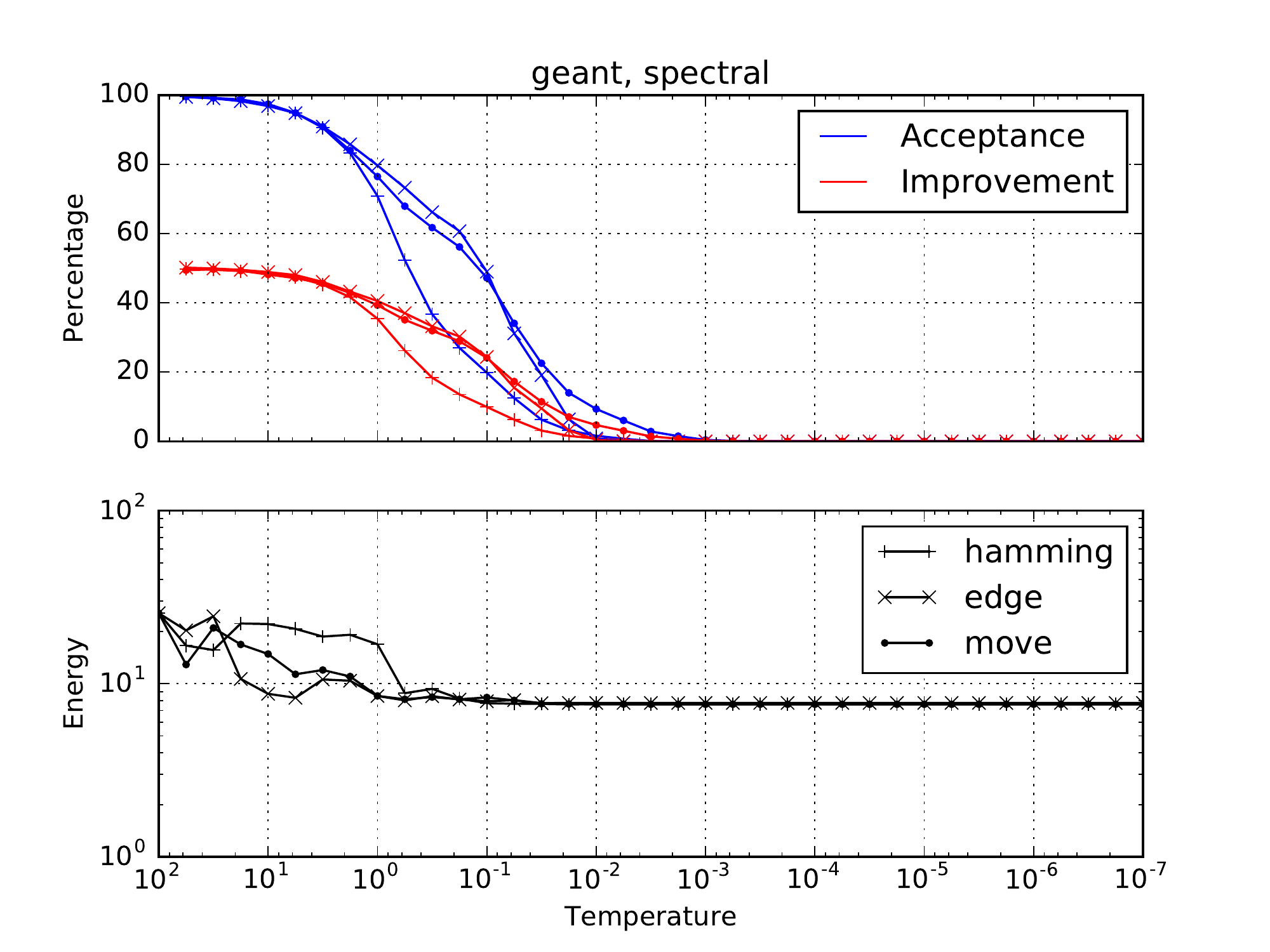}
      \caption{}
      \label{fig:geant-spectral-timeline}
   \end{subfigure}
   \begin{subfigure}[b]{0.49\linewidth}
      \includegraphics[width=\columnwidth]{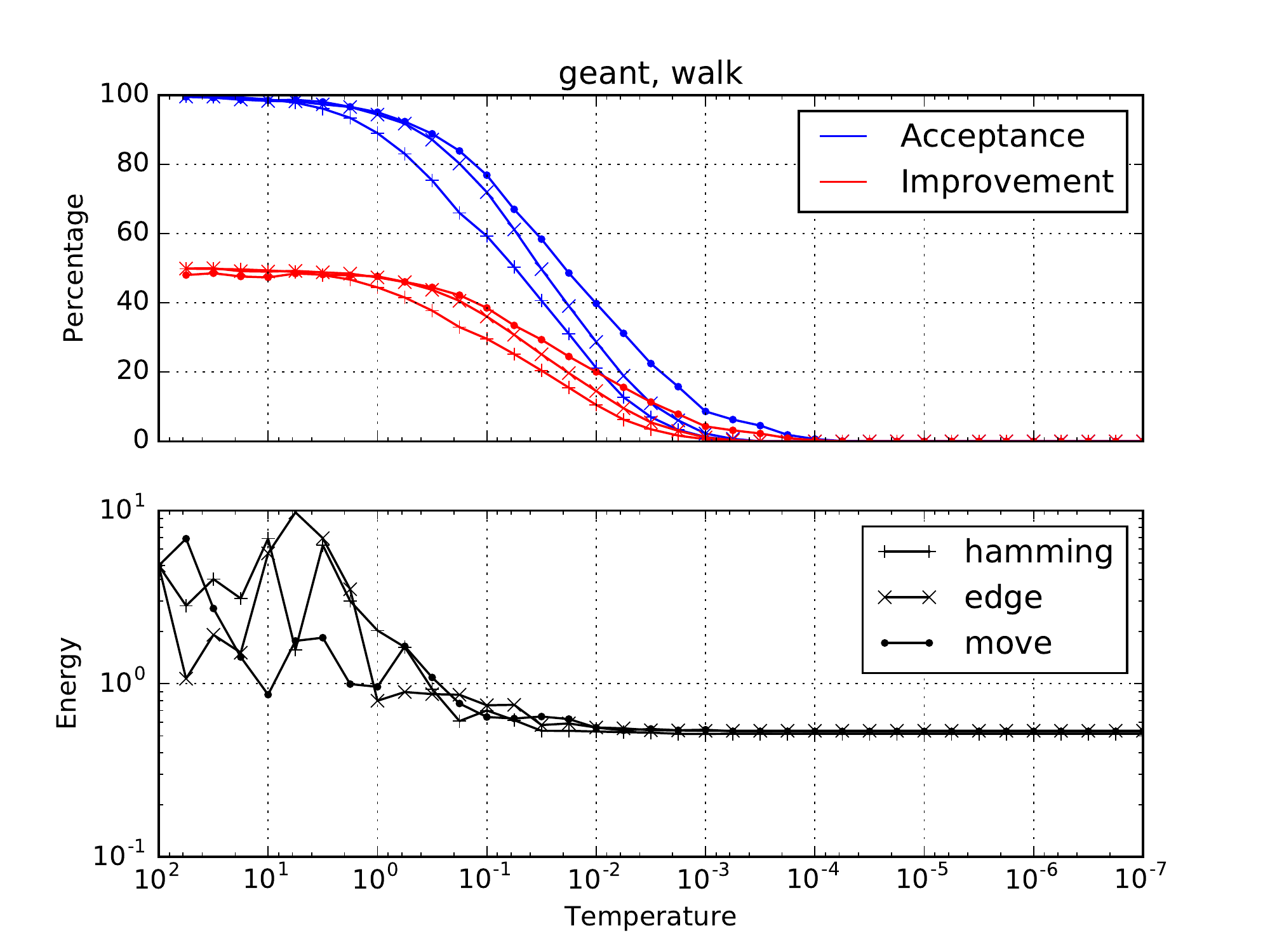}
      \caption{}
      \label{fig:geant-walk-timeline}
   \end{subfigure}
   \caption{
      Annealing timeline for different perturbation schemes when optimising
      \subref{fig:geant-spectral-timeline} spectral and
      \subref{fig:geant-walk-timeline} walk matrix distance,
      for graph `geant'.}
   \label{fig:geant-timeline}
\end{figure}

It is also helpful to compare the annealing timeline under different conditions.
These are shown in Figure~\ref{fig:geant-timeline} for the graph `geant'.
For the range of temperatures considered, annealing starts with a 100\%
acceptance rate and 50\% improvement rate, indicating that the algorithm is
correctly exploring neighbourhoods indiscriminately.
The acceptance rate starts to drop as the temperature is decreased, with a
corresponding drop in improvement rate, until both reach zero.
At lower temperatures, the algorithm explores less and favours only changes
towards the local minimum.
This behaviour is reflected in the energy values sampled at different
temperatures, which exhibit large variation at higher temperatures and
converge towards the final low energy state at lower temperatures.
Timelines for the other graphs exhibit a similar pattern.

\subsection{Comparison of indices for similar graphs}
\label{sec:results:indices}

Finally we compare the values of a number of indices on graphs for the best
output using each graph distance function with those of the original graph.
These are tabulated in Table~\ref{tab:indices} for graphs G12, G14, `geant',
`abstract', and `benzenoid'.
Note that we do not include `honey' because the original graph is not connected,
so that many of these indices cannot be computed.

\inserttabd{tab:indices}{cc|cc|c|ccc|ccc}
{Comparison of indices for original graphs and similar nested split graphs
obtained with simulated annealing. The closest nested split graph for a given
index is marked (*).}{
\multicolumn{2}{c|}{\emph{Graph}} & \multicolumn{2}{c|}{\emph{Edge Count}} & \emph{Degree} & \multicolumn{3}{c|}{\emph{Distance}} & \multicolumn{3}{c}{\emph{Eigenvalues}} \\
&& \emph{Edges} & \emph{Entropy} & \emph{Randić} & \emph{Wiener} & \emph{Szeged} & \emph{Co-PI} & \emph{Estrada} & \emph{Gutman} & \emph{Resolvent} \\
\hline
\multirow{3}{*}{\rotatebox[origin=c]{90}{G12}}
& original & 13  & 3.700  & 4.982  & 186  & 200  & 88   & 36.16  & 12.66  & 1.016 \\
& walk     & 28  & 4.807  & 5.010* & 104  & 186* & 158  & 406.1  & 15.97  & 1.051 \\
& spectral & 15* & 3.907* & 4.087  & 117* & 121  & 106* & 63.36* & 11.09* & 1.021* \\
\hline
\multirow{3}{*}{\rotatebox[origin=c]{90}{G14}}
& original & 18  & 4.170  & 5.698  & 231  & 399  & 136  & 59.3   & 14.71  & 1.014 \\
& walk     & 40  & 5.322  & 5.920* & 142  & 284* & 244  & 1473   & 20.26  & 1.049 \\
& spectral & 21* & 4.392* & 4.803  & 161* & 175  & 154* & 136.9* & 14.08* & 1.019* \\
\hline
\multirow{3}{*}{\rotatebox[origin=c]{90}{geant}}
& original & 64  & 6.000  & 17.51  & 4346  & 9503  & 1558  & 269.7             & 53.71  & 1.002 \\
& walk     & 213 & 7.735  & 14.59* & 1427  & 4395* & 4182  & $1.589\times10^7$ & 58.18* & 1.009 \\
& spectral & 76* & 6.248* & 9.912  & 1564* & 1770  & 1694* & 7068*             & 31.68  & 1.002* \\
\hline
\multirow{3}{*}{\rotatebox[origin=c]{90}{abstract}}
& original & 605  & 9.241  & 28.56  & 3062  & 10063  & 7772   & $1.436\times10^{10}$  & 126    & 1.009 \\
& walk     & 455* & 8.830* & 21.14* & 2967  & 14147  & 13692  & $3.793\times10^{10}$* & 87.13* & 1.006* \\
& spectral & 443  & 8.791  & 20.6   & 2979* & 10629* & 10186* & $6.797\times10^{10}$  & 79.33  & 1.006* \\
\hline
\multirow{3}{*}{\rotatebox[origin=c]{90}{benzenoid}}
& original & 129  & 7.011  & 47.80  & 149789 & 141852  & 8180  & 284.2                & 143.1  & 1 \\
& walk     & 4297 & 12.07  & 47.83* & 4823   & 40841*  & 36544 & $1.708\times10^{39}$ & 197.2* & 1.157 \\
& spectral & 122* & 6.931* & 13.21  & 8998*  & 9160    & 9038* & $3.205\times10^{4}$* & 35.91  & 1* \\
}

From these results we can observe the following general patterns.
Limiting NSGs optimized for spectral distance tend to have a number of edges
that is closer to that of the original graph than those optimized for walk distance.
Consequently, indices that directly depend on edge count (such as entropy) are
better approximated by these graphs.
These graphs are also better approximations for indices that depend directly on
the eigenvalues, as is to be expected.
On the other hand, graphs optimized for walk distance are better approximations
when computing indices that depend on the distribution of vertex degrees,
such as the Randić index.
In fact, the approximation here is often particularly close.
Finally, for indices that depend on the distribution of distances between
vertex pairs, the better approximator depends on the specific index.
The Wiener index and Co-PI index are better approximated by graphs optimized
for spectral distance, while those optimised for walk distance are better
for the Szeged index.


\section{Conclusions}
\label{sec:closure}

In this paper we have presented a method based on simulated annealing to
obtain a NSG that approximates a given real complex graph.
Three different perturbations have been considered for traversing the space of
NSGs on a given number of vertices, and two distance measures to determine how well a
specific NSG approximates the original graph.
Practical results for six graphs from social networks, communication networks,
word associations, and molecular chemistry, show that the simulated annealing
process is stable and converges to suitable NSGs.

We have also developed very efficient algorithms to compute a number of useful
graph indices, taking advantage of the geometrical properties of NSGs.
For the same six graphs, we have compared the indices of the original graphs with
those of the corresponding approximate NSGs, using each of the two distance
functions.
We see that the ideal distance function to use depends on the indices to
be computed.

The results shown in this paper are for relatively small graphs, with
at most 100 vertices and 1000 edges.
For practical purposes it would be interesting to see how well the methods
presented here scale to much larger graphs, such as those from the
Stanford Large Network Dataset, with $10$--$10\,000\times$ more vertices
and edges.
The efficient algorithms presented here to compute the graph indices scale
well to graphs of that size, because their complexity is polynomial in the
number of cells of the NSG, rather than polynomial in the number of vertices.
This is a consequence of the structural properties of the NSG.
However, the simulated annealing method as presented here does not scale
so well.
Optimizing the simulated annealing method is possible, and we leave this
for future work.



\section*{References}
\bibliographystyle{elsarticle-num}
\bibliography{references}

\end{document}